# A Hardware Friendly Unsupervised Memristive Neural Network with Weight Sharing Mechanism

*Zhiri Tang[a], Ruohua Zhu[a], Peng Lin[a], Jin He[a], Hao Wang[a], Qijun Huang[a], Sheng Chang[a*], Qiming Ma[b*]*

[a] *School of Physics and Technology, Wuhan University, Wuhan, Hubei Province, 430072, China*
[b] *Institute of Electrical Engineering of Chinese Academy of Sciences, Beijing, 100190, China*

*Abstract*—Memristive neural networks (MNNs), which use memristors as neurons or synapses, have become a hot research topic recently. However, most memristors are not compatible with mainstream integrated circuit technology and their stabilities in large-scale are not very well so far. In this paper, a hardware friendly MNN circuit is introduced, in which the memristive characteristics are implemented by digital integrated circuit. Through this method, spike timing dependent plasticity (STDP) and unsupervised learning are realized. A weight sharing mechanism is proposed to bridge the gap of network scale and hardware resource. Experiment results show the hardware resource is significantly saved with it, maintaining good recognition accuracy and high speed. Moreover, the tendency of resource increase is slower than the expansion of network scale, which infers our method's potential on large scale neuromorphic network's realization.

*Index Terms*—memristive neural networks (MNNs), digital integrated circuit, spike timing-dependent plasticity (STDP), unsupervised learning, weight sharing mechanism

## 1. Introduction

Memristor was postulated by L.O. Chua [1] in 1971 as the fourth basic circuit element after resistor (R), inductor (L) and capacitor (C) in electrical circuits. In 2008, the HP Labs used $TiO_2$ to realize a practical memristor device [2] for the first time. The value of memristor depends on the amount of electricity flowing through it, so memristor can mimic the memory function of biological synapses. Recently, memristive neural networks (MNNs), which use memristors as the parts of the synaptic weight storage or the synaptic weight value updating, have become a hot research topic, for their potential on a lot of combinatorial descriptions of biological synapse's characteristics, such as spike timing-dependent plasticity (STDP), long-term potentiation (LTP), long-term depression (LTD) [3-10] and so on.

MNNs have immense potential applications in many areas including image recognition, neuron modeling and high performance computing [11-14]. However, memristor has its limitations as a novel device. Still now, most memristors' fabrications [15-19] are not compatible with mainstream integrated circuit technique and have large fluctuations in parameters [20-22], which limit their applications in real tasks. Under these technique limitations, realizing memristive characteristics by mature circuit methods [23] becomes a practicable way. Specially, standard integrated circuit technologies [24,25], such as digital signal processing (DSP), field programmable gate arrays (FPGA) and application specific integrated circuits (ASIC), have advantages like high operation speed, good noise robustness and great expansibility, which are convenient for memristive characteristics' implementations [26]. Likewise, realizing neural networks by mature hardware is also an achievable way for extremely high operation speed and relative low cost [27-29]. For example, TrueNorth-the neuromorphic hardware based on spike-timing prototyped by IBM, has quite low energy consumption for communication [30].

Some state-of-the-art works about MNNs, such as developing an unsupervised network structure with analog memristive synapses [31] and using memristive neural network to build hidden hyperchaotic attractor [32], were also reported. However, complex calculations are still needed in neural networks. How to mimic biological learning mechanism with memristor model by a simply method is still an open area. Originally, memristor model should be merged into neural networks algorithm naturally and bring out its intrinsic virtues, such as the relationship between the memristor value and the current through memristor. To realize it, novel MNN algorithms should be developed.

What' more, the efficiency of hardware implementation is another theoretical difficulty. Some current researches about carrying on MNNs on hardware do not take the hardware resources that the memristor model will consume into account. Y.V. Pershin and M. Di Ventra built a classic memristor emulator which realized all required synaptic properties [33] and designed memory circuit elements to mimic STDP efficiently after few years [34]. Both of these outstanding hardware memristor models can realize all characteristics of memristor including STDP. However, these models are so complex that they consume too many hardware resources and are too slow in practice. In other researches using FPGA to realize MNNs [24,25], hardware resources are obviously insufficient to support these large scale MNNs, because the computational complexity and the resources consumed increase severely with network's scale. Thus, it is quite vital to develop techniques to raise the efficiency of implementing memristor models and MNN algorithms for very-large-scale MNNs on hardware.

Inspired by the above, this paper designs a MNN algorithm suiting for hardware and studies its applications in image recognition. The main contributions can be summarized as follows:

* Corresponding author
[a] E-mail: changsheng@whu.edu.cn (Sheng Chang)
[b] E-mail: maqiming@mail.iee.ac.cn (Qiming Ma)

a. A novel unsupervised MNN algorithm is designed. It combines memristive intrinsic characteristics with learning algorithm smoothly by converting the spike time information into memristor value directly. Through this way, complex calculation is avoided and biological learning mechanism is mimicked in a simply method.
b. A weight sharing mechanism is proposed. It can reduce MNNs' hardware occupancies quite efficiently without performance degradation on training speed and classification accuracy, verified by testing with network scale expansion. It is very advantageous for large scale MNN hardware design.

## 2. Memristive Neural Network Algorithm

In this section, a hardware optimized algorithm is designed aiming to the high efficiency of integrated circuit. That includes the modified HP model and a new hardware friendly unsupervised spike response model.

### 2.1. Modified HP Model

In traditional HP model, relationship of the voltage at the two ends of the memristor (U(t)) and its current I(t) is

$$U(t) = [R_{off} - (R_{off} - R_{on})\mu_v \frac{R_{on}}{D^2} \int_{-\infty}^{t} I(t)dt] I(t) \quad (1)$$

When I(t) is fixed to 1, the above formula is simplified as:

$$U(t) = R_{off} - (R_{off} - R_{on})\mu_v \frac{R_{on}}{D^2} \int_{-\infty}^{t} dt = k_1 - k_2 \Delta t \quad (2)$$

where $k_1 = R_{off}$ and $k_2 = (R_{off} - R_{on})\mu_v \frac{R_{on}}{D^2}$. In this case, the memristor value can be simplified as:

$$M(t) = \frac{U(t)}{I(t)} = k_1 - k_2 \Delta t \quad (3)$$

Further, if the traditional HP memristor model is added with window function, it can be derived as follow:

$$W(t) = \int_{-\infty}^{t} \mu_v \frac{R_{on}}{D} I(t) f(x) dt = \mu_v \frac{R_{on}}{D} q(t) f(x) \quad (4)$$

So that the memristor value is:

$$M(t) = R_{on}(\mu_v \frac{R_{on}}{D^2} f(x) \int_{-\infty}^{t} I(t)dt) + R_{off}(1 - \mu_v \frac{R_{on}}{D^2} f(x) \int_{-\infty}^{t} I(t)dt)$$
$$= R_{off} - (R_{off} - R_{on})\mu_v \frac{R_{on}}{D^2} f(x) \int_{-\infty}^{t} I(t)dt \quad (5)$$

From above, it also can be transformed into Eq. (2). The complexity of the memristor model is independent of our simplified approach because what we use is the basic characteristic of memristor, which means our approach will make sense as long as the memristor value has linear relationship with the sum of current through memristor. That is friendly for hardware.

Based on this modified HP model, the time information in spiking neural networks, the time intervals between two spikes, can be described with the following structure in Fig. 1.

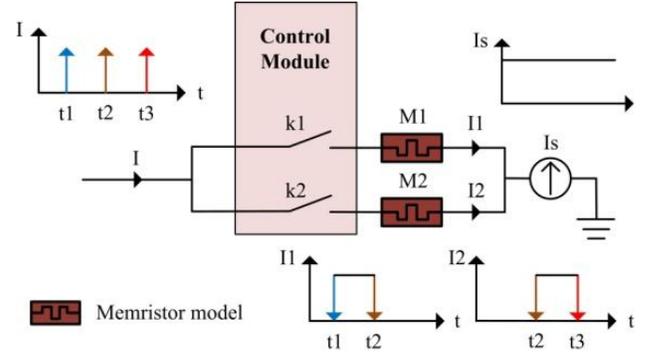

**Fig. 1** Time information transformation with the modified HP model.

In Fig. 1, the height of the spike signals is set to 1 unit and three continuous spikes, for an instance, fire in t1, t2 and t3. The one ends of the memristors are fixed to the constant 1 standard circuit, Is.

When the first spike arrives at t1, the control module closes the switch k1. When the second spike comes at t2, the switch k1 is disconnected and the switch k2 is closed. When the third spike comes at t3, switch k2 is also disconnected. According to the currents actually passing through the two memristors (I1 and I2), the value stored in the memristor M1 has a linear relationship with the time intervals between the first and the second spikes, and the value of memristor M2 has the same relationship with the time intervals between the second and the third spikes.

Current researches indicate that the rules for changing the biological synaptic weights follow the STDP [35], which means that the strength of connection between neurons in biology can be adjusted by the chronological order of reaching the synaptic signal [36], which is shown as follow:

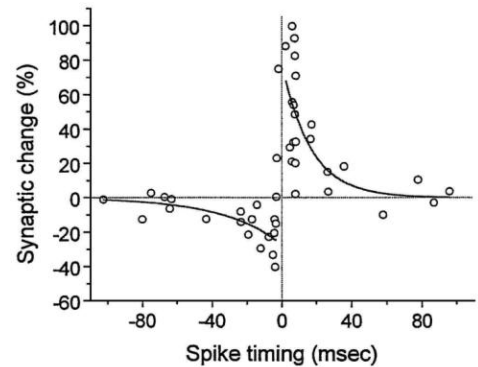

**Fig.2** Spike timing-dependent plasticity function.

The spike timing is firing time intervals between postsynaptic neurons and presynaptic neurons. The synaptic change (%) is the change of connection strength between neurons, as the synaptic weight change, dW, in neural networks.

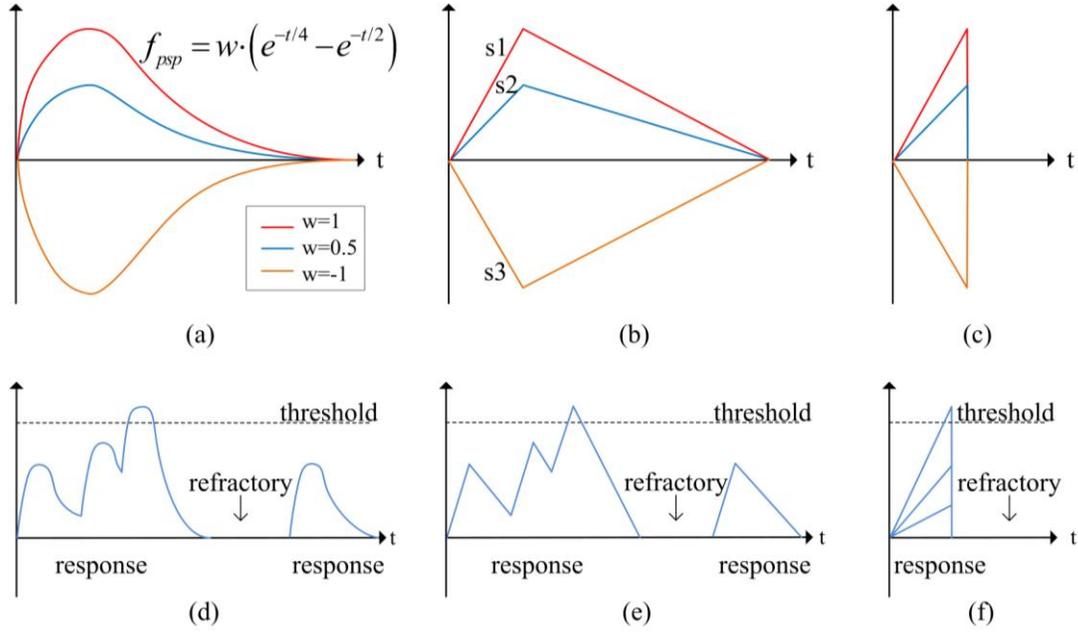

**Fig. 3** Simplification of post synaptic potential (PSP) and spike response model (SRM).

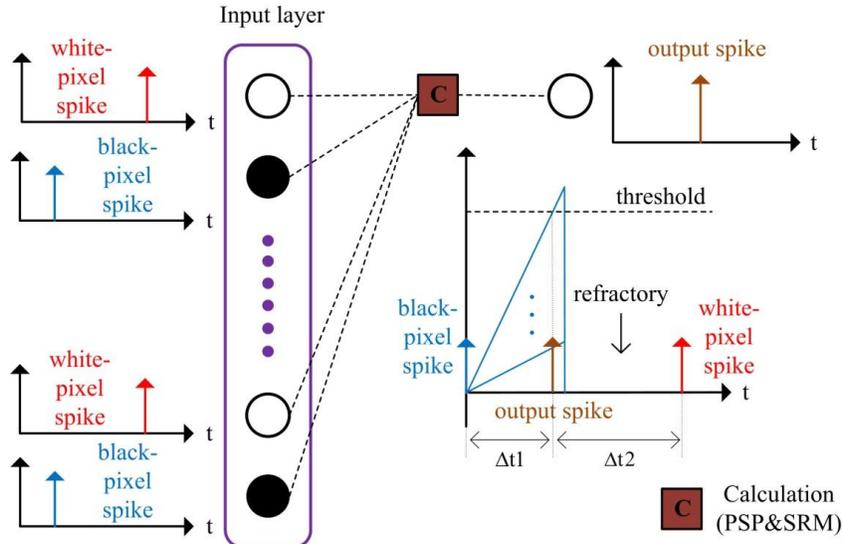

**Fig. 4** Basic network structure with simplified post synaptic potential (PSP) and spike response model (SRM).

In the process of realizing STDP using memristor model, the ordinate is dG-the change in the reciprocal of memristor value. So it has dW = dG. The dG of memristor represents the value of synaptic weight updating. dG is greater means dW is greater. Hence, the convergence speed is faster.

From above, the biologic plasticity can be described as follows:

$$dW = dG = \Delta(\frac{1}{k_1 - k_2 \Delta t}) \tag{6}$$

When the change of time interval between output spike and target spike-$\Delta(\Delta t)$-is positive, dG is positive. If $\Delta(\Delta t)$ is negative, dG is negative. The spike timing-dependent plasticity (STDP) is realized.

### 2.2. Unsupervised Spike Response Model

Spike response model (SRM) is chosen to develop our unsupervised mechanism due to its similarity to biological neurons [37-40]. Supposed that two neurons are connected and communicate with spikes, the transmitting neuron is called as the presynaptic neuron while the receiving one is called as the postsynaptic neuron. The membrane potential's activity is the main characteristic of spiking neurons which will stay at a resting value in refractory period.

When the spikes generated by the presynaptic neurons arrive to the postsynaptic neurons, it will contribute a post synaptic potential (PSP) to the membrane potential. A general PSP function is defined as:

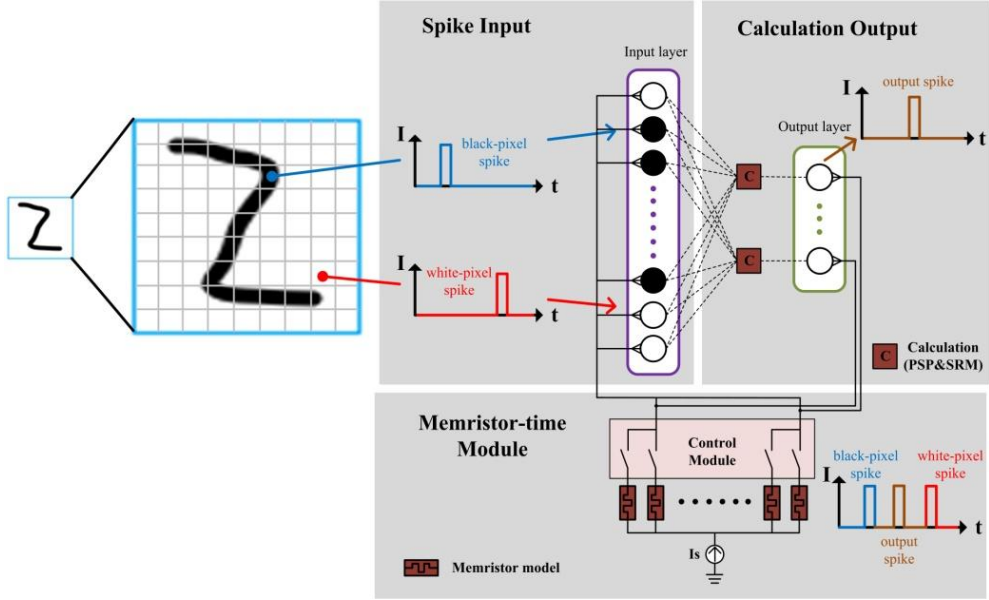

**Fig. 5** Hardware architecture of MNN.

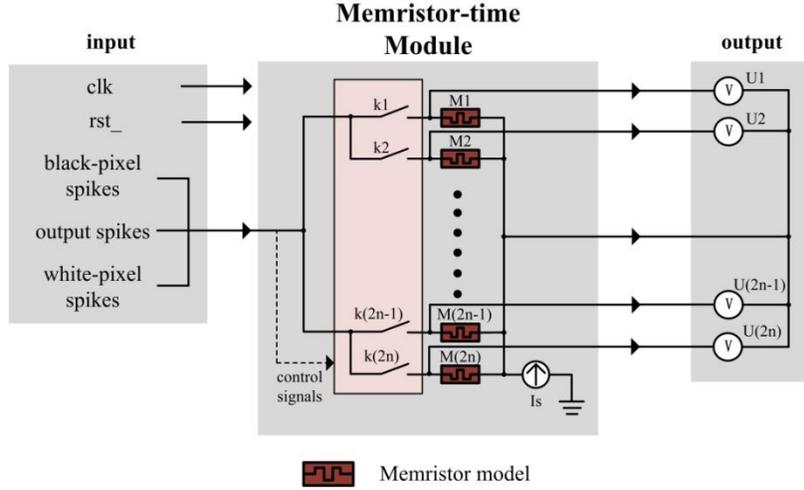

**Fig. 6** Circuit diagram of memristor-time module.

$$f_{psp}(t) = w(e^{-\frac{t}{4}} - e^{-\frac{t}{2}})$$
(7)

Fig. 3(a) gives its demonstration. When the addition of PSP generated by presynaptic neurons on the membrane potential is up to the threshold value in the response period, the postsynaptic neurons will fire and go into the refractory period. Until the refractory period ends, the postsynaptic neurons will return to the response period as shown in Fig. 3(d).

For the purpose of hardware friendly implementation, we linearize Eq. (7) as Fig. 3(b). Since the most important information is the peak and rise time of PSP, the linearization process directly connects the origin and the maximum points, which means it is similar to the slope of positive edge of PSP. Thought calculation, the linearization equation is as follow and $a=0.09$, $b=2.77$:

$$f_{psp(lin)}(t) = a \cdot w \cdot t \quad (t \leq b)$$
(8)

The slopes s1, s2 and s3 are 1, 0.5 and -1 respectively after the linearization. Fig. 3(d) is linearized as Fig. 3(e).

To further simplify this model, the presynaptic neurons are assumed to fire at the same time and the threshold $v_T$ is set to a small value. In this case, the presynaptic neuron's firing will trigger the postsynaptic neuron's firing and the falling edge of the firing can be ignored. So Fig. 3(c) and (f) are got. What's more, if the time intervals between this firing and the next one are long enough, the refractory period after the firing don't need to be considered.

Based on the above, the neural network training process with the simplified PSP and SRM is developed as shown in Fig. 4. The binary images are set as the black-pixel spikes and the white-pixel spikes. The output neurons can fire as output spikes. As the instance of Fig. 1, there are three spikes from these three types of neurons. Setting a reasonable threshold, the postsynaptic neurons can fire between the black-pixel and white-pixel spikes. One can see:

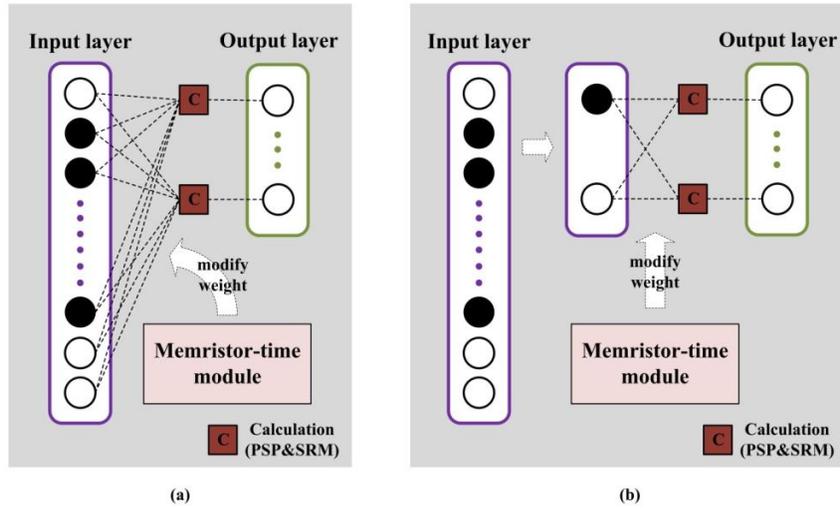

**Fig. 7** (a) MNNs without weight sharing mechanism. (b) MNNs with weight sharing mechanism.

$$\Delta t_1 = t_{output} - t_{black} > 0$$
$$\Delta t_2 = t_{output} - t_{white} < 0$$
(9)

Take Eq. (9) into Eq. (6), so we have:
$$dW_{black} = dG_{black} > 0$$
$$dW_{white} = dG_{white} < 0$$
(10)

$dW_{black}$ is the synaptic weight updating of black neuron and $dW_{white}$ is the synaptic weight updating of white neuron. $dG_{black}$ is the change in derivative of memristor value of black neuron and $dG_{white}$ is the change in derivative of memristor value of white neuron.

The relationship between input and output neurons is the combination of PSP and SRM. Input neurons will cause changes in voltage based on PSP and the threshold in SRM determines the fire time of output neurons. For example, if all the training images are divided into three categories, there will be three output neurons and three output spikes. Due to STDP, the closer the two spikes distances are, the larger the corresponding synaptic weight is. In other words, the correct classification, whose synaptic weight is increased to the largest, means that the fire time of the output neuron is closer to the black spike. If the output is expected as the first category, the first output neuron should fire first due to the increased black-pixel synaptic weights, while the expected firing time of other two categories will be delayed due to the decreased black-pixel synaptic weights.

Through such process, the output spikes are described as self-competition. If the first output neuron fires when the test images are input, the first category is the classification result. This simplified unsupervised algorithm is convenient to be operated on hardware.

## 3. Network Design with Weight Sharing Mechanism

Hardware design methodology for MNNs is introduced in this section, especially the sharing of synaptic weight and its update mechanism.

### 3.1. Hardware Architecture of Network

The overall network is divided into three major modules: spike input module, calculation output module and memristor-time module. First of all, the input images are coded into spikes as mentioned above. A black-and-white binary image is divided into two kinds of spikes, and the neurons corresponding to the black-pixel spikes fire at the same time. After a certain period of time, the neurons corresponding to the white-pixel spikes will fire. The encoded information are sent into the input layer consisted of spike input modules. In the output layer, the data are processed by the calculation output modules, including the modified PSP and SRM rules. We estimate the approximate fire time of output spikes and set the white spikes in the refractory period. So that the black-pixel spikes, output spikes, and white-pixel spikes are got in turn.

According to our algorithm, inputting these three spikes into the memristor-time module can directly obtain two synaptic weight updating values. Each output spike corresponds to two memristor models. The first records the time intervals between the black-pixel spikes and the output spikes, and the other records the time intervals between the white-pixel spikes and the output spikes. This process is controlled by an internal control module for each switch. The different memristor units can record data for different periods of time and convert them into the updated synaptic weights. Then bring the updated synaptic weights into training and continue this cycle until the end of the training. The total hardware architecture is shown as Fig. 5.

Specially, the circuit diagram of memristor-time module is given as Fig. 6. It converts the time interval information of spikes into the values of memristors. The inputs to this module are black-pixel spikes, white-pixel spikes and output spikes, and these spikes serve as signals for controlling all switches at the same time. Setting the number of classifications as n, the number of switches, memristors and outputs are all 2n. Since the current is set to 1, the output voltage across a memristor equals to the value of the memristor.

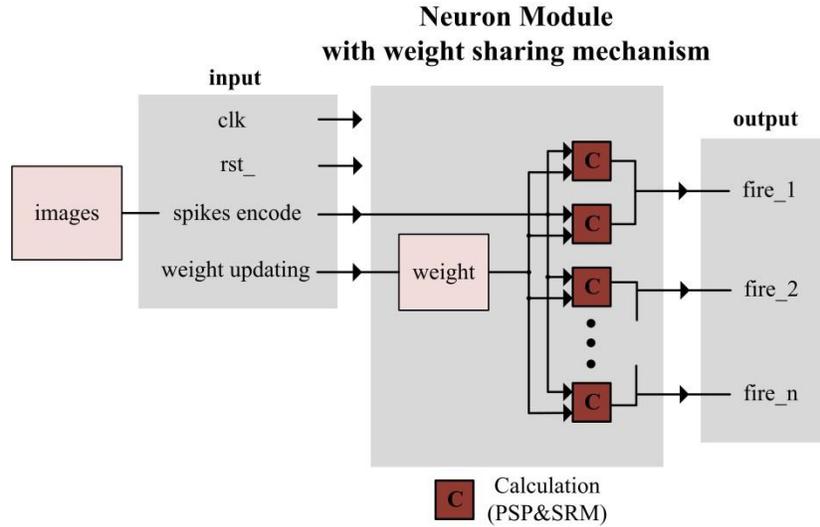

**Fig. 8** Circuit diagram of neuron module with weight sharing mechanism.

*3.2. Memristive Weight Sharing Mechanism*

As one of the main issues in hardware implementation of neural networks, resource occupancy is always been focused on. As the connection between neurons, the number of synapsis (synaptic weights) sharply increases with the enlarging of network. In order to reduce resource occupancy and allow hardware architecture to accommodate very-large-scale networks, we propose a new weight sharing mechanism. From the above discussion of algorithm, one can see the synaptic weight updating is calculated and stored by the memristor-time module. Using this sharing mechanism, the number of synaptic weights can be significantly reduced, which is hardware friendly.

From Fig. 5 one can see, the firing time of the black spike is the same. So if the initial values of the synaptic weights corresponding to the black pixels are set as the same, the amount of change obtained by the memristor-time module is also the same. So the synaptic weights of all black pixels can be directly set to a single value which is efficient to save hardware resource.

The case of the corresponding synaptic weights for the white pixels is alike. Using the weight sharing mechanism, the number of synaptic weights can be reduced to just double of the number of categories, as shown in Fig. 7. For instance, if the trained and tested images are 3*3 pixels and the number of categories is 3 (that's a 3*3*3 network), then both of the required synaptic weights and the number of updated memristive synaptic weights are six under our weight sharing mechanism.

The circuit diagram of neuron module with weight sharing mechanism is shown as Fig. 8. The inputs to this module are clock, reset, spikes encode from training images and synaptic weight updating from memristor-time module, arranging all the process of adding, calculation and output with timing control. The synaptic weight updating is added to the old synaptic weight firstly, and then the new synaptic weight calculates with the spikes encoding to get output fire. The number of classifications in a figure is n, so that the number of fires is n. The output of neuron module is the main input of memristor-time module and the output of memristor-time module is the input of neuron module except spikes encoding.

**4. Experimental Results and Analysis**

In this section, the performances of our MNN circuit and weight sharing mechanism, such as the recognition accuracy, the resource cost and the maximum frequency, are discussed.

*4.1. Pattern Recognition Results*

A classic alphabet recognition database is used to test our hardware's performance. For example, Fig. 9(a) gives the case of three alphabets, 3*3 'ZVN', which is suitable for the test of three classifications. As the result, the 3*3*3 network by our method can deal with this problem efficiently. The total training is completed in three cycles and all of the images in Fig. 9(b) can be correctly classified.

The case of a 9*9 'ZVNXC' dataset is selected to detailedly analyze our method's accuracy in a larger scale as Fig. 10(a). Fig. 10(b) and (c) show the training processes of the networks without and with the weight sharing mechanism. In them, the abscissas are time, and the ordinates are the values of synaptic weight updating. A training or test cycle takes 20us including five different images which are Z, V, N, X and C in turn. One can see that in all cases, when all the dWs drop below value of the set threshold, the training process comes to end. We set the threshold for dWs only instead of specific training cycle number. With the weight sharing mechanism, the black pixels and the white pixels correspond to their certain synaptic weights, respectively. Therefore, the values of synaptic weight updating calculated for each category are the same in Fig.10(c) while those are different in Fig. 10(b). Noticeably, the training process can be completed in three cycles with weight sharing mechanism, which is the same as without weight sharing mechanism. So the introduction of weight sharing mechanism has no influence on the training speed. That is to say, it can reduce resource occupancy on hardware without affecting training efficiency.

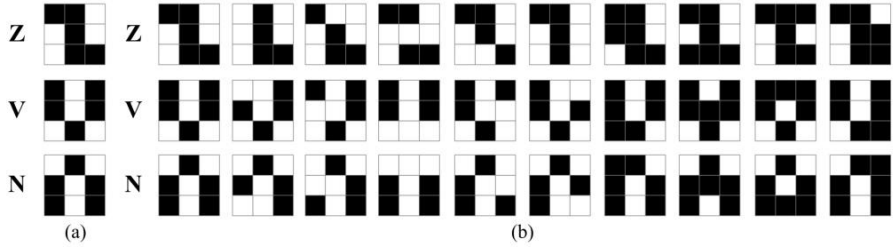

**Fig. 9** The 3*3database for (a) training and (b) testing.

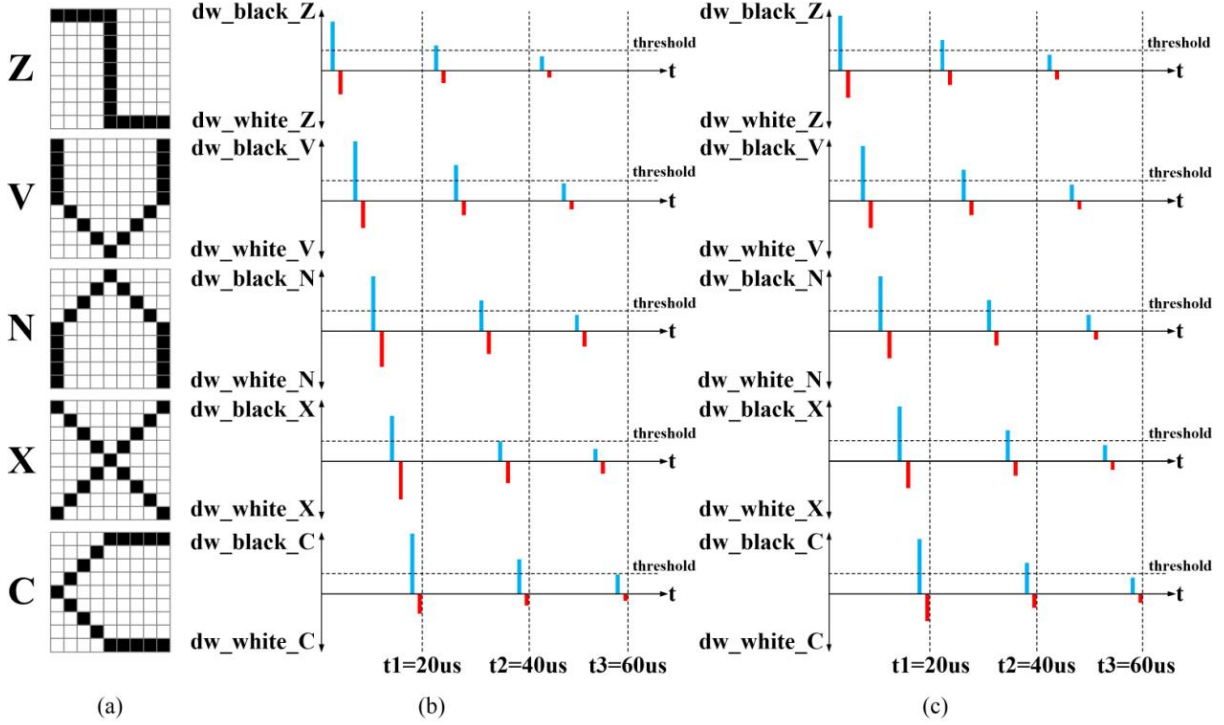

**Fig. 10** (a) The 9*9*5 network training images. (b) The changes of memristors' values in training process without weight share mechanism. (c) The changes of memristors' values in training process with weight share mechanism.

During the test, we add random noise to each of these five types of standard images, i.e. randomly changing some pixels in Fig. 10(a). 150 sets of images - a total of 750 images - are tested. The test results are shown as Table 1. From the table, one can see only 7 images are identified into the wrong classifications among the 750 images. The correct rate is over 99%.

**Table 1**
Test results of the 9*9*5 network

|  |  | Expectation |  |  |  |  |
|---|---|---|---|---|---|---|
|  |  | Z | V | N | X | C |
| Experiment results | Z | 149 | 0 | 0 | 1 | 0 |
|  | V | 0 | 149 | 0 | 0 | 1 |
|  | N | 1 | 0 | 148 | 0 | 1 |
|  | X | 0 | 1 | 0 | 149 | 0 |
|  | C | 0 | 1 | 1 | 0 | 148 |

### 4.2. Resource Occupancy

In order to verify the optimization of the hardware resource occupancy by this weight sharing mechanism, we select six networks from small to large to carry out experiments on Stratix V(5SGXEA7N2F45C2), which are the 3*3*3 network, the 5*5*3 network, the 7*7*3 network, the 5*5*5 network, the 7*7*5 network and the 9*9*5 network. Besides that, to verify platform's influence on our method's resource cost, we test the 3*3*3 network as an example on three different-level FPGAs, Stratix V: 5SGXEA7N2F45C2, Cyclone 10 LP: 10CL120YF780I7G and Arria 10: 10AX115U1F45I1SG. These two experimental results are listed as Table 2 and Table 3.

From the tables, it's clear that the use of weight sharing mechanism reduces the hardware resource occupancy significantly. The relationship between the increase in resource occupancy and the number of categories is greater than the correlation with the number of input neurons. Defining the number of input neurons as neuron$^2$ ($N^2$) and the number of classifications as type (T), we can set the abscissa as $N*T^2$ and the ordinate as the resource footprint to draw a resource occupancy tendency as Fig. 11, in which the black dashed line represents a linear function as a reference.

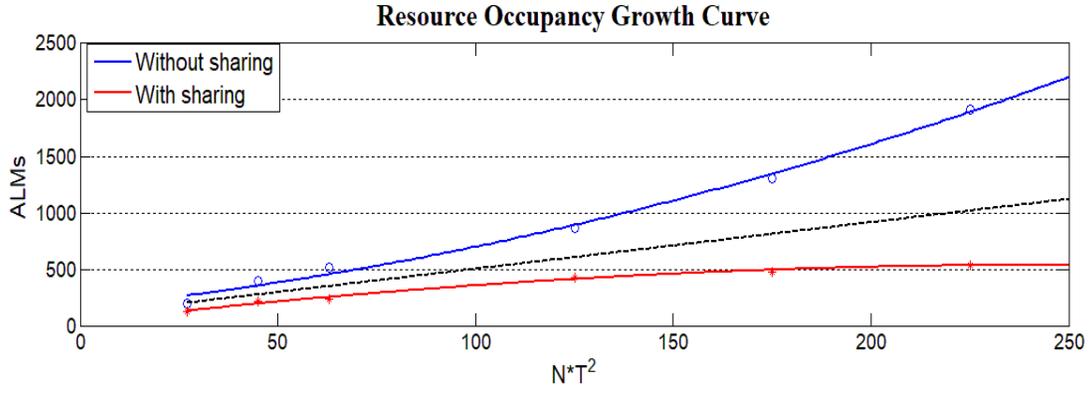

**Fig. 11** Resource occupancy growth curve.

Table 2
Resource occupancies of different scale networks on Stratix V

| Network scale | Without weight sharing (in ALMs) | With weight sharing (in ALMs) |
|---|---|---|
| 3*3*3 | 199 | 122 |
| 5*5*3 | 395 | 214 |
| 7*7*3 | 516 | 237 |
| 5*5*5 | 869 | 428 |
| 7*7*5 | 1309 | 475 |
| 9*9*5 | 1917 | 540 |

Table 3
Resource occupancies of the 3*3*3 network on different hardware platforms

| Hardware platforms | Without weight sharing | With weight sharing |
|---|---|---|
| Stratix V(in ALMs) | 199 | 122 |
| Cyclone 10(LE) | 542 | 477 |
| Arria 10(in ALMs) | 203 | 130 |

Table 4
Maximum clock frequencies of different scale networks without weight sharing mechanism

| Fmax(MHz) | Stratix V | Cyclone 10 | Arria 10 |
|---|---|---|---|
| 9*9*5 | 247.52 | 128.3 | 235.02 |
| 7*7*5 | 243.07 | 125.06 | 226.55 |
| 5*5*5 | 256.54 | 126.5 | 227.32 |
| 7*7*3 | 248.51 | 126.21 | 242.01 |
| 5*5*3 | 250.69 | 125.42 | 237.25 |
| 3*3*3 | 270.12 | 126.29 | 246.67 |

Table 5
Maximum clock frequencies of different scale networks with weight sharing mechanism

| Fmax(MHz) | Stratix V | Cyclone 10 | Arria 10 |
|---|---|---|---|
| 9*9*5 | 249.88 | 140.11 | 248.76 |
| 7*7*5 | 258.93 | 137.78 | 245.16 |
| 5*5*5 | 267.52 | 138.5 | 240.21 |
| 7*7*3 | 242.84 | 137.87 | 244.74 |
| 5*5*3 | 258.2 | 136.87 | 246.37 |
| 3*3*3 | 250.25 | 138.08 | 244.98 |

Table 6
Comparisons of resource occupancy and clock frequency

| Design | Original model[25] (two-compartment) | Modified model[25] (two-compartment) | Our method | Our method |
|---|---|---|---|---|
| Platform | Cyclone IV: EP4CE115 | Cyclone IV: EP4CE115 | Cyclone IV: EP4CE115 | Stratix V: 5SGXEA7N2F45 |
| Recourse | 3250 (LE) | 3031 (LE) | 135 (LE) | 70 (in ALMs) |
| Mem-bits | 122880 | 0 | 0 | 0 |
| Multiplier | 288 | 0 | 0 | 0 |
| PLLs | 0 | 0 | 0 | 0 |
| Fmax | 17.66MHz | 40.68MHz | 86.21MHz | 181.16MHz |

From the fitting curves in the figure, one can see that our weight sharing mechanism has an obvious effect of reducing hardware resource cost. Without this mechanism, the increase of resource occupancy is faster than the expansion of network scale while it is quite slower after adding the weight sharing mechanism. The network is larger, weight sharing's advantage is more significant. It can be inferred that our mechanism has great potential facing the implementation of very-large-scale networks on hardware.

*4.3. Maximum Clock Frequency*

Since hardware's parallelism calculation structure is consistent with neural networks, the operating speed of neural networks on hardware is much faster than that on software. Therefore, we analyze the maximum clock frequencies of different network scales.

To avoid platform's influence, three different-level FPGAs, Stratix V, Cyclone 10 LP and Arria 10, are all tested. The networks' sizes are selected as 3*3*3, 5*5*3, 7*7*3, 5*5*5, 7*7*5 and 9*9*5. What's more, pipeline design and clock constraints are both used in all designs to optimize the maximum clock frequencies. The experimental results are listed in Table 4 and Table 5.

Table 4 shows the maximum clock frequencies of different scale networks without weight sharing mechanism while Table 5 gives the results with weight sharing mechanism. It is clear that with the expansion of network scale, the maximum clock frequency shows a downward trend, but the decline is very slow and the frequencies are stable at a high level. This phenomenon is clear in all three different FPGAs, which identifies our mechanism's universality. Moreover, comparing the results in Table 5 with the data in Table 4, one can see that in most cases the maximum clock frequency is improved, although the increase is not too much. Since the introduction of the weight sharing mechanism will not change the learning latency number as shown in Fig. 10, this frequency improvement verifies that the weight sharing mechanism can efficiently save the resource cost and improve the operating speed at the same time.

*4.4. Performance Comparison*

To evaluate our hardware method's performance, comparison with other work has been done. Although in this cutting-edge area, comparable researches from algorithm improvement to hardware optimization are not enough, a state-of-the-art work result [25] and our method are listed in Table 6. In the test, the same network scale which includes 10 classification numbers is used. To ensure the test's equity, our method's platform is also degraded to Cyclone IV.

From the results one can see, our method has significant advantages on both resource occupancy and operation speed in the same platform. If changing to the mainstream platform (Stratix V), our method's performance can be improved further.

5. Conclusions

In this paper, a hardware friendly unsupervised algorithm for memristive neural networks is presented. Neuron and network structures with digital integrated circuit are developed and a weight sharing mechanism is proposed, which bridge the gap between large network scale and hardware resource. The experiments in different-level FPGAs and different network scales show that our network structure and sharing mechanism not only reduce resource cost significantly but also have a nice trend with the expansion of network scale, maintaining good recognition accuracy and high operating speed. We hope these ideas can give an inspiration for the design of memristive neural networks and other neuromorphic networks.

**Acknowledgements**

This work was supported by the National Natural Science Foundation of China (61574102, 61774113 and 61331007), the Fundamental Research Fund for the Central Universities, Wuhan University (2042017gf0052 and 2042016kf0189) and the Natural Science Foundation of Hubei Province, China (2017CFB660).